\documentclass[journal]{IEEEtran}
\usepackage{cite}
\ifCLASSINFOpdf
\else
\fi
\usepackage{epsfig}
\usepackage{subfigure}
\usepackage{verbatim}
\usepackage{amsfonts}
\usepackage{amssymb}
\usepackage{amsmath}
\usepackage{cuted}
\usepackage{enumerate}
\usepackage{graphicx}
\usepackage{multirow}
\usepackage{makecell}

\usepackage{float}
\usepackage[table,xcdraw]{xcolor}

\usepackage{color}
\usepackage{soul}
\usepackage{footnote}

\makesavenoteenv{table}

\newcommand{\AzCom}[1]{}
\newcommand{\AzDel}[1]{}

\newcommand{\boris}[1]{\textcolor{blue}{[#1]}}

\begin{document}
%

\title{HARE: Supporting efficient uplink multi-hop communications in self-organizing LPWANs}

\author{\IEEEauthorblockN{Toni Adame, Sergio Barrachina, Boris Bellalta, Albert Bel}
\IEEEauthorblockA{\\Department of Information and Communication Technologies, \\ Universitat Pompeu Fabra, Barcelona\\
Email: \{toni.adame, sergio.barrachina, boris.bellalta, albert.bel\}@upf.edu}
}


\maketitle

\begin{abstract}
The emergence of low-power wide area networks (LPWANs) as a new agent in the Internet of Things (IoT) will result in the incorporation into the digital world of low-automated processes from a wide variety of sectors. The single-hop conception of typical LPWAN deployments, though simple and robust, overlooks the self-organization capabilities of network devices, suffers from lack of scalability in crowded scenarios, and pays little attention to energy consumption. 

Aimed to take the most out of devices' capabilities, the HARE protocol stack is proposed in this paper as a new LPWAN technology flexible enough to adopt uplink multi-hop communications when proving energetically more efficient. In this way, results from a real testbed show energy savings of up to 15\% when using a multi-hop approach while keeping the same network reliability. System's self-organizing capability and resilience have been also validated after performing numerous iterations of the association mechanism and deliberately switching off network devices.
\end{abstract}



\section{Introduction} \label{introduction}
In the coming years, electronic equipment will be interconnected and consequently every person and every industry will become simultaneously data generators and consumers. Internet of Things (IoT) paradigm is a key enabler of this vision by delivering machine-to-machine (M2M) and machine-to-person communications on a massive scale.

As more and more things are connected to the Internet, low-cost and low-traffic devices are starting to be demanded. However, traditional cellular networks do not deliver a good combination of technical features and operational cost for those IoT applications that need wide-area coverage combined with relatively low bandwidth, long battery life, low hardware and operating cost, and high connection density \cite{jones2016top}. 

\begin{figure}
\centering
\includegraphics[width=9cm]{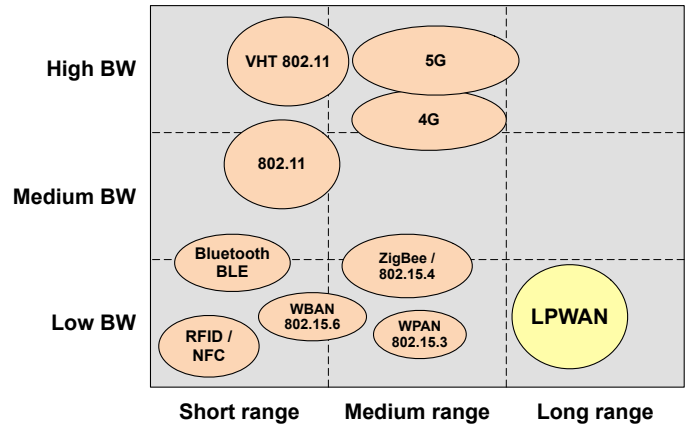}
\caption{Localization of LPWAN technologies according to range capability and bandwidth required.}
\label{fig:lpwan_map}
\end{figure}

Low-power wide area networks (LPWANs) are intended to become the engine of long-range, low-bandwidth IoT applications (see Figure \ref{fig:lpwan_map}), which until now have been constrained by deployment costs and power issues. The goal of these networks is to deliver small amounts of data over long ranges, at rates of up to tens of kilobits per second (kbps), with a battery lifetime of up to several years, supporting thousands of devices connected to a base station, and facilitating online integration. 

Existing LPWAN technologies can be categorized into two types \cite{markkanen2015lpwa}:
\begin{enumerate}
\item \textit{Dedicated LPWANs} consisting of the purposely designed technologies such as LoRa\texttrademark \cite{alliance2016lora}, SIGFOX\texttrademark \cite{sigfox2016main}, Ingenu\texttrademark \cite{ingenu2016main}, Weightless\texttrademark \cite{weightless2016setting}, DASH7 \cite{dash72016main}, and ETSI-LTN \cite{etsiltn2016main}.
\item \textit{Evolutionary LPWANs} covering the alternatives that have been developed as upgrades to well-established protocols like IEEE 802.11ah (also known as Wi-Fi HaLow) \cite{adame2014ieee}, EC-GSM-IoT \cite{ecgsmiot2016gsma}, LTE-M \cite{ltem2016gsma}, and NB-IoT \cite{nbiot2016gsma}.
\end{enumerate}

LPWAN architecture is characteristically single-hop, where end devices are connected directly to the base station, greatly simplifying the network and endowing it with robustness and centralized control. And yet this single-hop massive channel access sets out some inherent challenges: reliability, scalability, flexibility, and quality of service (QoS). In fact, the channel access mechanism of some LPWAN technologies resorts to the use of ALOHA \cite{goursaud2015dedicated} \cite{laya2016goodbye}, a random medium access control (MAC) protocol in which end devices transmit without doing any carrier sensing to check the channel state in advance. Although simple, this uncontrolled medium access leads to interference or packet collisions among uncoordinated devices, acutely affecting reliability in dense networks. In addition, LPWAN devices located far away from the base station must make use of high transmission power levels, resulting in severe energy consumption and reduced battery lifetime \cite{barrachina2016multi}.

In this article, the HARE protocol stack is proposed as a new LPWAN technology flexible enough to adopt uplink multi-hop communications when proving energetically more efficient than single-hop. A full set of advanced techniques belonging to different communication layers has been designed for this purpose, while ensuring data transmission reliability: 

\begin{itemize}
\item Inherent clock synchronization, with nodes being periodically set in time by means of beacons.
\item TDMA-like channel access for groups of contenders with multiple transmission opportunities.
\item Adaptive transmission power level.
\item Flexible and scalable network association process.
\item Energy-aware, adaptive and resilient routing protocol.
\item Regular use of deep-sleep states.
\end{itemize}

Furthermore, HARE protocol stack has been implemented and tested in real hardware platforms. Results evaluation from different network configurations (single-hop vs. multi-hop, use of different MAC layers within the TDMA slots, channel error injection) show very high reliability while maintaining low energy consumption (particularly in multi-hop topologies). Lastly, we have observed a better overall system behaviour when using multi-hop topologies in non error-prone scenarios.

The remainder of this paper is organized as follows: Section \ref{scenario} introduces the main requirements of feasible use cases for HARE. Next, Section \ref{operation} describes the general operation of the protocol stack and Section \ref{protocol} provides detailed information of the developed mechanisms. Section \ref{testbed} describes the proposed testbed and Section \ref{results} compiles the obtained results from different experiments. Lastly, Section \ref{conclusions} presents the conclusions and discusses open challenges.

\section{Scenarios and Requirements}
\label{scenario}
According to their own characteristic range and bandwidth capabilities with respect to other technologies, the main use cases to which LPWANs are addressed include security alarms, car park spaces, agricultural applications, smart metering, consumer electronics, and intelligent buildings. By way of illustration, in the ENTOMATIC EU-project\footnote{ENTOMATIC main webpage: https://entomatic.upf.edu/} a network of wireless sensor nodes \cite{potamitis2017automated} periodically report information on pest population density and environmental parameters, such as temperature and relative humidity.

HARE is clearly aligned with typical LPWAN applications and circumscribes its suitability to those scenarios with special concern for energy efficiency, where device batteries are so limited that the establishment over time of a direct connection to the base station, or gateway (GW), would greatly affect their lifetime. 

In this sense, Table \ref{requirements} offers a comprehensive list of common requirements from use cases to which HARE protocol stack gives response in combination with the appropriate hardware. Assuming that this hardware provides good signal penetration, a single GW can serve up to thousand devices within its given coverage range. Applications executed by stations (STAs), in turn, follow a \textit{continuous data delivery model} \cite{chen2004qos} for their sensed information, periodically delivering small amounts of non-delay sensitive data.

\begin{table}[]
\centering
\caption{Common requirements of HARE use cases}
\label{requirements}
\begin{tabular}{|l|l|}
\hline
\textbf{Requirement} & \textbf{Value}                                                   \\ \hline
Coverage range       & Up to several km.                                                \\ \hline
Geographic coverage  & Excellent even in remote and rural areas                \\ \hline
Penetration          & Good in-building and in-ground penetration                       \\ \hline
\begin{tabular}[l]{@{}l@{}}Device density\\ (per base station)\end{tabular}     & High (up to thousand) \\ \hline
Power profile        & Unassisted, battery-powered devices \\ \hline
Battery lifetime & From some months up to several years
\\ \hline
Throughput           & \textless 100 bits/s                                             \\ \hline
Latency              & Non-delay sensitive                                              \\ \hline
Mobility             & Static devices                                                   \\ \hline
Cost                 & Low hardware and operating cost                                  \\ \hline
Maintenance & Unassisted and self-organizing network\\
\hline
Delivery model & Continuous data delivery model\footnote{Although not considered in the current article, future HARE developments will consider offering QoS in scenarios with miscellaneous sensors (continuous, event-driven, query-driven, and hybrid)}\\
\hline
\end{tabular}
\end{table}

As sensor nodes are scattered over large areas, sometimes with problematic access, self-maintenance of the system shall be a priority, capable of giving response to the following challenges:

\begin{enumerate}
\item Node initiated network connection: Once installed for the first time or relocated, any node shall initiate its association process through a simple action (for instance, pressing a button). 
\item Self-configuration and management: With the aim of building a robust network, it shall adapt itself to environmental and/or topology changes without human intervention. 
\item Battery lifetime maximization: LPWANs replace old monitoring systems consisting of assigning human resources to study in situ the behavior of one or more physical parameters. Therefore, maximizing battery lifetime in such systems is vital in order to justify their usage ahead of other methods.
\item Firmware distribution: Any change in the network configuration or in the application purpose shall be remotely and easily distributed by the GW.
\end{enumerate}

Lastly, the operating system of embedded sensor nodes is typically less complex than general-purpose operating systems \cite{farooq2011operating}. However, the high variety of resources to manage in this kind of devices (processors, memories, clocks, network interfaces, etc.) and the demand of support for concurrent execution of processes (time synchronization, data acquisition, task scheduling, channel access, routing parameters, etc.) make essential the use of a real time operating system (RTOS). 

Under these premises, Table \ref{apps} compiles some of the main use cases supported by the HARE protocol stack in five IoT representative sectors: home and industrial automation, public infrastructure, natural resources, and smart agriculture and farming.

\begin{table}[]
\centering
\caption{Use cases supported by HARE protocol stack}
\label{apps}
\begin{tabular}{|l|l|}
\hline
\textbf{Sector}                                                                     & \textbf{Use cases}                                                                                                                                                                                \\ \hline
\textbf{Home automation}                                                            & \begin{tabular}[c]{@{}l@{}}Domotics\\ Child/elderly tracking\\ Smart metering\end{tabular}                                                                                                                    \\ \hline
\textbf{Industrial automation}                                                      & \begin{tabular}[c]{@{}l@{}}Remote maintenance/control\\ Logistics \\
Local asset tracking management\\ Distribution automation (smart grid)\\ Smart metering\end{tabular} \\ \hline
\textbf{Public infrastructure}                                                      & \begin{tabular}[c]{@{}l@{}}City smart lighting\\ Smart parking\\ Intelligent buildings\\ Predictive maintenance\end{tabular}                                                                              \\ \hline
\textbf{Natural resources} & \begin{tabular}[c]{@{}l@{}}Environmental monitoring\\ Natural disasters detection\end{tabular}           \\ \hline
\textbf{\begin{tabular}[c]{@{}l@{}}Smart agriculture\\ and farming\end{tabular}} & \begin{tabular}[c]{@{}l@{}}Agriculture monitoring\\ Animal monitoring\\ Silo stock monitoring\end{tabular}                                                          \\ \hline
\end{tabular}
\end{table}



\section{HARE operation} \label{operation}

The HARE protocol stack conceives end devices as elements controlled by the GW by means of beacons. This centralized approach allows STAs to remain asleep the majority of the time, so that their single concern is to be awake enough in advance to listen to the next beacon. Network synchronization is thus achieved and allows the GW to ask for specific data and/or distribute configuration changes in just one hop.

The GW is considered to be appropriately placed close to a power source. Thus it may always stay in an active state and is provided with the ability to directly communicate (i.e., via single-hop communications) with any node of the network through \textit{unicast} and/or \textit{broadcast} messages as well as to redirect gathered data from the WSN to other networks or the Internet. 

\begin{figure}
\centering
\subfigure[Multi-hop LPWAN with a gateway (GW) and 30 stations ($N_{1}-N_{30}$) deployed in 4 rings ($R_{1}-R_{4}$).]
{\includegraphics[width=8cm]{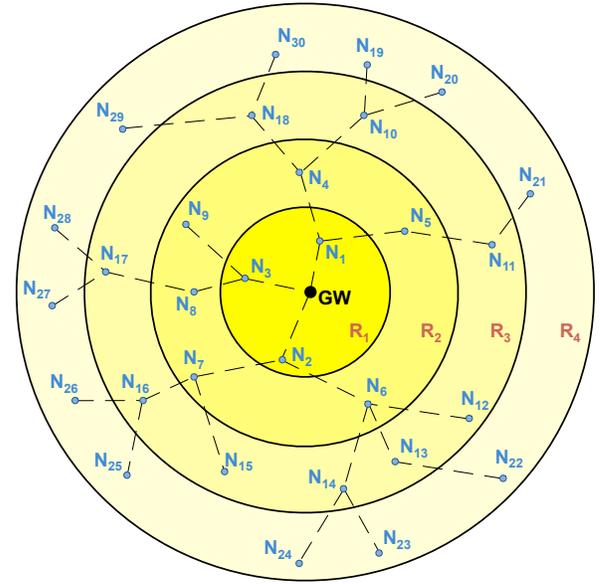}}
\subfigure[Multi-hop LPWAN affected by an obstacle, with a gateway (GW) and 10 stations ($N_{1}-N_{10}$) deployed in 4 rings ($R_{1}-R_{4}$).]
{\includegraphics[width=8cm]{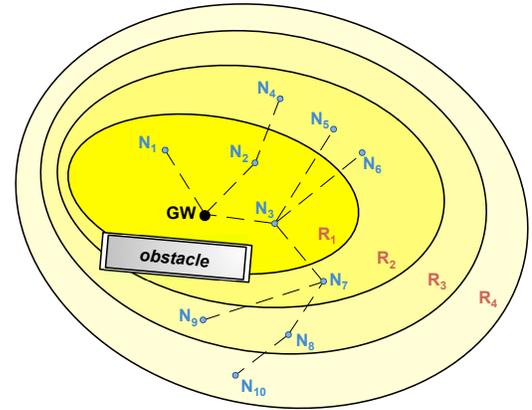}}
\caption{Network topology of typical LPWANs}
\label{fig:top}
\end{figure}

Conversely, STAs can take advantage of their neighbors to create multi-hop paths over which data is transmitted to the GW by means of lower transmission power levels. Depending on their position within these paths, STAs are ideally organized into \textit{rings}, as shown in Figure \ref{fig:top}. The number of hops to reach the GW determine the ring number (i.e., STAs from \textit{ring 2} need two hops to reach the GW). 

Each uplink data transmission phase (consisting of one or more  \textit{transmission windows}) begins with a beacon signal from the GW. Transmission windows are in turn virtually split into as many TDMA slots as network rings, so that STAs are only active during their own slot (for transmitting data) and the previous one belonging to their children\footnote{\textit{Children} refers to all STAs of an adjacent higher ring from which an STA receives packets. Similarly, \textit{parent} refers to that STA from an adjacent lower ring to which an STA transmits its own packets (after aggregating the ones from its children) in its way to the GW.} (for receiving data).

The first slot is allocated to the highest ring and the rest are scheduled consecutively. Data received by STAs is aggregated to that generated by themselves, and finally sent to the corresponding parent at the minimum power level which ensures reliable communications. This process is repeated as many times as rings the network has.

The correct reception of data transmissions at the GW is acknowledged with a broadcast message, so that STAs are not only aware of their own end-to-end reliability, but also of those STAs in the same path to the GW. These acknowledgment beacons, together with the information obtained from their adjacent nodes, allow STAs to decide whether they should remain awake to perform retransmissions of lost network packets.

Network association (also started by a beacon) remains stable until a change in the topology is detected or the mechanism is reset by the GW. Nevertheless, the agreed transmission power between adjacent nodes in the association phase is constantly monitored and adjusted in a decentralized way in order to reduce the energy consumption.


\section{Protocol stack} \label{protocol}
The main features of the HARE protocol stack are shown in Table \ref{stack}; a complete description of them is offered next.

\begin{table}[h]
\centering
\caption{\label{stack} Main features of HARE protocol stack}
\begin{tabular}{|c|c|}
\hline 
\textbf{Layer} & \textbf{Features} \\ 
\hline 
\textbf{Transport} & \makecell{End-to-end ACK\\ Poisoning mechanism \\ Transmission windows \\ Distributed caching} \\ 
\hline 
\textbf{Network} & \makecell{Addressing system\\ Association \\ Routing} \\ 
\hline 
\textbf{Link} & \makecell{Beaconing system\\ Wakeup patterns \\ Data transmission, aggregation \& segmentation \\ Power regulation mechanism}  \\ 
\hline 
\rowcolor[HTML]{EFEFEF}
\textbf{Physical} & \textit{Hardware dependant} \\ 
\hline 
\end{tabular} 
\end{table}

\subsection{PHY layer}
HARE protocol stack is intended to be used over any wireless PHY layer fulfilling a minimum set of functions; namely, availability of different operational states both in the microprocessor (processing, low power mode) and in the radio module (receiving, transmitting, and sleeping), selection of different transmission levels in the radio transceiver, and ability to execute low level tasks required by typical shared medium access techniques.

\subsection{Link layer}
The MAC layer is a combination of a time division multiple access (TDMA) scheme, where time slot duration is managed by the GW, and an underlying carrier sense multiple access with collision avoidance (CSMA/CA) technique with packet acknowledgment (ACK), performed by the group of STAs allocated into each generated time slot. At this point it is worth noting that HARE is not only limited to CSMA-based access techniques, but also can properly work with other MAC protocols for WSNs \cite{cano2011low}.

\subsubsection{Beaconing system}
\begin{figure*} [t!!]
\centering
\includegraphics[width=16cm]{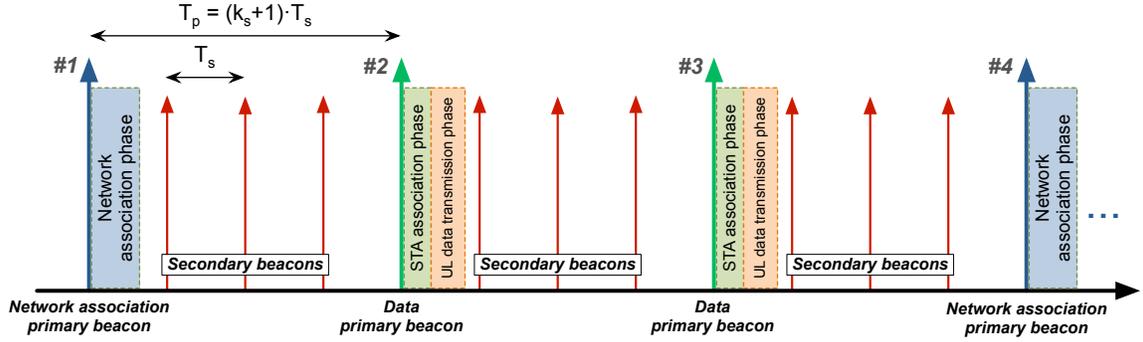}
\caption{HARE beaconing system consisting of \textit{network association primary beacons}, \textit{data primary beacons}, and \textit{secondary beacons}.}
\label{fig:beaconing}
\end{figure*}

The designed beaconing system has a double function: synchronizing the network devices and scheduling the different actions to be performed. Two types of beacons are used for this purpose: \textit{primary} and \textit{secondary beacons} (see Figure \ref{fig:beaconing}).

Both beacons include a timestamp, the time until the next \textit{primary beacon}, and the next action to be taken by the network: for instance, an association phase (\textit{network association primary beacon}), or an uplink data transmission phase (\textit{data primary beacon}). \textit{Secondary beacons} include the same information as \textit{primary} ones, and are used to guarantee information redundancy for already associated STAs as well as to accelerate network discovery for non-associated ones. However, no action is performed by STAs after a \textit{secondary beacon}.

Time between two consecutive \textit{primary beacons} and two consecutive \textit{secondary beacons} is defined as $T_{p}$ and $T_{s}$, respectively. Where $T_{p} = (k_{s}+1) \cdot T_{s}$, being $k_{s}$ the number of \textit{secondary beacons} transmitted after every \textit{primary beacon}.

\subsubsection{Wakeup patterns}
A wakeup pattern is a set of instructions generated by the GW which define the wakeup plan of its associated STAs over time periods. With the goal of minimizing the time STAs remain active (and, consequently, their energy consumption), two different wakeup patterns controlled by the GW are proposed according to the network's traffic flow \cite{keshavarzian2006wakeup}.

\begin{figure} [h!] 
\centering
\includegraphics[width=7cm]{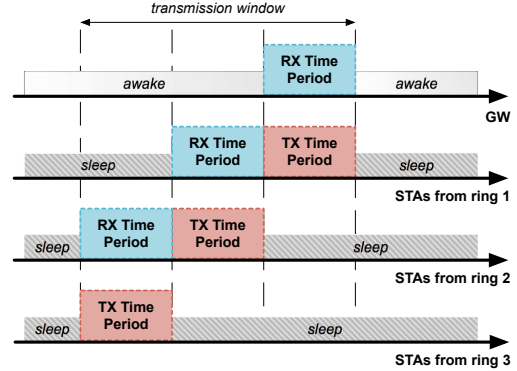}
\caption{Example of a \textit{staggered wakeup pattern} in a 3-ring LPWAN performing uplink communications.}
\label{fig:staggered}
\end{figure}

The \textit{periodic wakeup pattern} is suitable for listening to broadcast downlink communications from the GW, as it makes all STAs wake up at the same time. On the other hand, uplink communications follow a \textit{staggered wakeup pattern}, which allocates different active periods to nodes belonging to adjacent rings with partial overlapping (as shown in Figure \ref{fig:staggered}). Apart from reducing time STAs are awake during uplink communications, this method facilitates the implementation of data aggregation mechanisms.

Even though STAs have predetermined active periods, they can go to sleep even earlier in the transmitting (TX) time period if their parent has acknowledged all their data, or in the receiving (RX) time period after having received all data from their children.

\subsubsection{Data transmission, aggregation, and segmentation}

Downlink communications are generally executed through broadcast messages from the GW. Conversely, uplink communications are unicast and follow a multi-hop route.

The \textit{staggered wakeup pattern} fits here perfectly with the approach of data aggregation in WSN. Thus nodes attach their own data to that received from their children and all the information is jointly sent to the next hop (i.e., parent). If the total amount of data aggregated by an STA exceeds the maximum payload supported by the hardware, it is split into segments\footnote{The amount of data aggregated by an STA (from itself and from its children) is called \textit{packet}. If this packet is split into different parts, each one of these parts is called \textit{segment}. In case both terms can be indistinctly used, the current article will use \textit{packet}.} sent consecutively. 

A selective ACK mechanism has been developed,
so that before the end of the allocated time slot, the receiver explicitly lists which segments in a stream coming from the same child are acknowledged. Upper layers are therefore responsible for making the sender retransmit only the missing segments in successive transmission windows. 

\subsubsection{Power regulation mechanism}

The selection of the minimum suitable transmission power level for outgoing packets is managed through a mechanism based on the received signal strength indicator (RSSI). For this purpose, a safety margin for reliable communications is defined by $\text{RSSI}_{\min}$ and $\text{RSSI}_{\max}$. If a node is transmitting data packets (ACKs) to its parent (child) at a power level making the received RSSI higher than $\text{RSSI}_{max}$, it will be asked to decrease it for the next transmission. Similarly, if the received RSSI is lower than $\text{RSSI}_{min}$, it will be asked to increase it. 

Power regulation requests are included in an RSSI control field of data packet and ACK headers. Possible values of this field are: \textit{increase}, \textit{keep}, and \textit{decrease}. Once computed the requests from parent and children, the STA determines whether and how to regulate its own power level depending on the following considerations:
\begin{itemize}
\item If one or more STAs ask for a higher value, increase the power level.  
\item If all STAs ask for a reduction, decrease the power level. 
\item Otherwise, keep the current power level.
\end{itemize}

In addition, if an STA needs to retransmit a packet to its parent, it will also increase the power level in each new transmission window. Regarding the association process, whenever an STA listens to a discovery request, it will answer at maximum power. The STA selected as parent will keep the maximum power level at the beginning and regulate it following the previously described procedure. Instead, those STAs not selected as parents will set their power back to the level they had before answering to the discovery request.

\vspace{5mm}
Consequently, the main advantages of using such a MAC layer scheme are:
\begin{itemize}
\item Clock synchronization is inherent to TDMA, with nodes being periodically set in time by means of beacons.
\item Groups of nodes have their time slots clearly allocated, and collisions within groups are sensibly reduced or even avoided by using CSMA/CA. 
\item Network overall lifetime is increased by putting nodes in non-active modes for most of the time and only periodically waking up to check for activity. 
\item Association and routing mechanisms are also fit for this scheme, so that intermediate and already associated nodes do not have to constantly listen to hypothetical network discovery requests.
\item The scheme is also suitable for uplink data aggregation.
\item Changes in the network configuration or even new firmware can be easily distributed in a coordinated manner.
\end{itemize}

\subsection{Network layer}
Network communications follow a centralized scheme, where the GW adopts the main role and assumes the responsibility of managing network associations, delivering network addresses, and periodically notifying the start of new routing processes.  

STAs adopt a subordinated role waiting for orders coming from the GW. In the routing process, they organize themselves in paths autonomously, but all subsequent data transmissions are addressed to the GW, directly or through other STAs.  Conversely, the GW can make use of its greater transmission power to periodically send broadcast messages to all network STAs, or send unicast messages to single STAs.

\subsubsection{Addressing system}

The addressing system is managed by the GW, which allocates a unique network address to each node during the association process. Nodes will maintain the same network address as long as they do not leave the network. A dynamic record matching the MAC and the network address of all STAs is stored in the GW. The size of the network address is configurable and its value determines the addressing range.

\subsubsection{Association}
To cope with multiple association requests in a short period of time, the system is able to admit new STAs through two different mechanisms: an active, global, scheduled one, called \textit{network association mechanism}; and a passive, singular one, called \textit{STA association mechanism}.

\begin{itemize}
\item \textbf{Network association mechanism} \newline
The \textit{network association mechanism} allows a large amount of STAs to associate to the network in a short period of time. Once the GW is activated, or after a pre-determined number of \textit{primary beacons} $(N_{\text{pr}})$, the GW broadcasts a \textit{network association primary beacon}.
\newline
Depending on the RSSI value received in the \textit{network association primary beacon} as well as some other configuration parameters, STAs determine their turn to initiate the association process (generally, the greater the RSSI received, the earlier association turn is selected). 
\newline
STAs then follow with a discovery message sent via broadcast, which is responded by the GW and all the already associated STAs, provided they are within the coverage range. The process of selecting the best path to reach the GW is detailed in the \textit{Routing} subsection.
\newline
Once the routing mechanism is completed, the GW notifies the joining of new STAs by means of a summary broadcast message sent immediately after every association turn. 

\item \textbf{STA association mechanism} \newline
The \textit{STA association mechanism} provides a solution to those specific nodes that (i) have not found a path to the GW during the network association mechanism, (ii) have been powered on between two consecutive \textit{network association primary beacons}, or (iii) have simply suffered routing problems in their path to the GW.

This mechanism follows the same pattern as the \textit{network association} one, with the single exception that there is only one association turn located immediately after each \textit{data primary beacon} to be used by non-associated STAs. 
\end{itemize}

Inactive or erratic STAs are removed from the network and the GW's routing table  to create, if necessary, new routing paths that ensure correct packet reception from remaining network STAs. Disassociations can be controlled by the GW through the \textit{disassociation mechanism} or by the STAs themselves through the \textit{self-disassociation mechanism}:

\begin{itemize}
\item \textbf{Disassociation mechanism} \newline
The GW removes an STA from the network if not receiving any data packet during a pre-determined number of consecutive \textit{primary beacons} $(N_{\text{pd}})$. A roster with the latest disassociated STAs is included in every \textit{primary beacon}. This information is not only useful for malfunctioning STAs, which can make immediate use of the \textit{STA association mechanism}, but also for their parents, as they can check the current state of their children. Hence, if all its children became disassociated, a parent would go to sleep during the RX time period allocated to its ring.




\item \textbf{Self-disassociation mechanism} \newline
The goal of this mechanism is to avoid repetitive association requests and other energy consuming procedures that could make STAs run out of battery when no connection with the GW is possible. All STAs have a timer that is activated after being switched on or when receiving a \textit{primary beacon}. From that moment on, if an STA does not receive any other beacon during a predetermined period $(T_{d})$, it turns itself off. Thus the STA is considered \textit{dead} and it will need to be reactivated by manual procedures.

\end{itemize}

\subsubsection{Routing}
\label{routing}
The routing protocol has been designed as an intrinsic part of the association process. Thus, according to the responses to the discovery message coming from other nodes, each STA determines which candidate is the best one to become its parent; i.e., the one with the minimum $S$ value from:
\begin{eqnarray}
S=a_{1} \cdot (P_{\text{TX}_{max}}-\text{RSSI}_{\text{TX}})+a_{2} \cdot (P_{\text{TX}_{max}}-\text{RSSI}_{\text{RX}})+a_{3} \cdot r+a_{4} \cdot c,
\label{routing_weights}
\end{eqnarray}
where $P_{\text{TX}_{max}}$ is the maximum transmission power of the transceiver (in dBm), $\text{RSSI}_{\text{TX}}$ is the RSSI received at the candidate (in dBm), $\text{RSSI}_{\text{RX}}$ is the RSSI received at the STA itself (in dBm), $r$ is the ring to which the candidate belongs, and $c$ is the current number of candidate's children. The $a$ weights are attached to every \textit{primary beacon}, and can be tuned by the GW according to environment requirements.

Once computed the best parent, the STA sends it a specific request. This request will be forwarded by the parent through its own path until reaching the GW, which will send a packet via broadcast confirming the association and providing the STA with its new address. This way, both the newly associated STA and its parent are informed of the establishment of the new path. 

When the association process is finished, the STA exactly knows the next hop its messages must follow to reach the GW. As long as the STA is associated to the network, it uses the same routing path, which is only recomputed after an internal or external (i.e., from its parent) failure. Indeed, no new routing process is initiated unless it is part of a new \textit{network association mechanism}.


\subsection{Transport layer}

Reliable end-to-end communications from the STAs to the GW, where retransmissions are only executed when needed and by the minimum number of involved STAs, are achieved in HARE by using the following mechanisms:


\begin{figure*}[t!!]
\centering
\includegraphics[width=16cm]{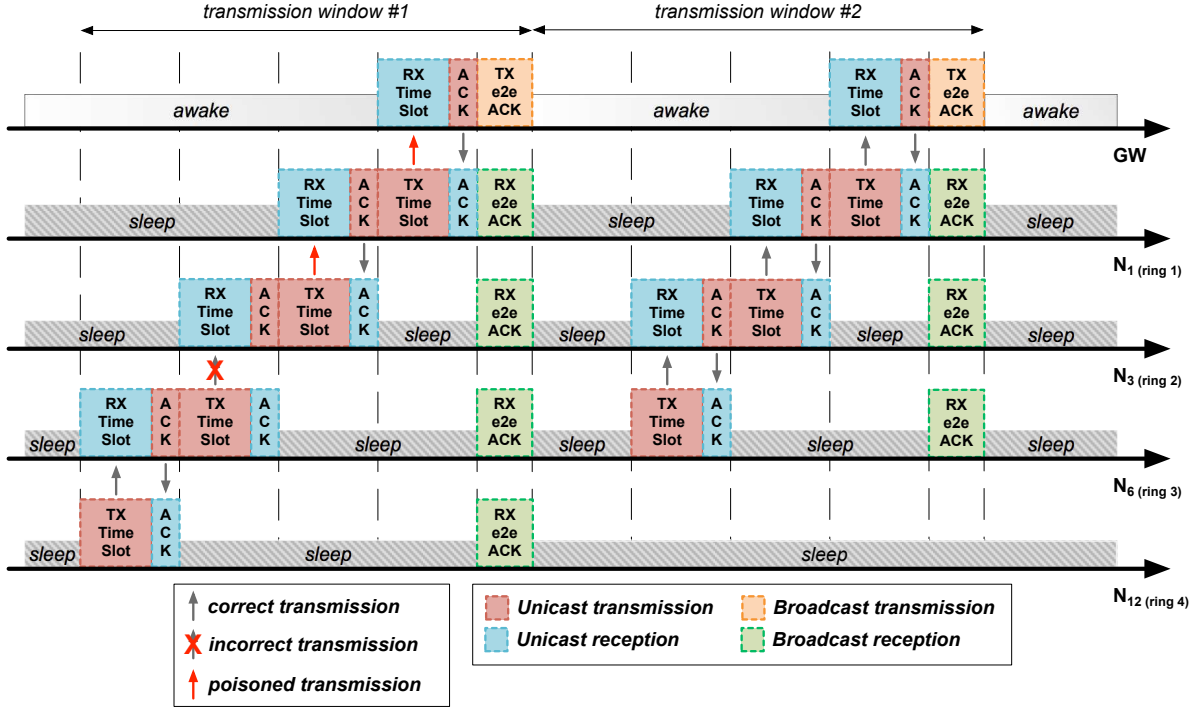}
\caption{Uplink data transmission phase in a multi-hop LPWAN running HARE protocol stack with the network topology from Figure \ref{fig:poison1}. Note the communication problems in the first transmission window between nodes $N_{6}$ and $N_{3}$.}
\label{fig:txwindow}
\end{figure*}

\subsubsection{End-to-end ACK}
According to the \textit{staggered wakeup pattern}, STAs from ring 1 are the last ones to access to the channel and transmit their information. Once compared the data sources with the expected uplink traffic, the GW emits a broadcast message called \textit{end-to-end ACK} (e2e ACK) with a list of acknowledged STAs. Figure \ref{fig:txwindow} shows the e2e ACK operation at the end of every transmission window. Apart from being simple, quick and simultaneously listened by all network elements, end-to-end ACKs allow STAs to evaluate the state of their path to the GW and act consequently.

\subsubsection{Poisoning mechanism}
The poisoning mechanism identifies which specific nodes experience communication problems in their path to the GW, so that they can perform subsequent retransmissions. Nodes having problems with their children transmit packets with the poison flag activated. An STA is considered \textit{poisoned} if, before transmitting an outgoing data packet, one of the following conditions is satisfied: 

\begin{itemize}
\item The STA is part of a poisoned path; i.e., it has received one or more packets with the poison flag activated during the current transmission window.
\item The STA has not received any data packet from one or more of its children.
\item The STA has not received all the expected segments from one or more of its children.
\end{itemize}

\begin{figure}
\centering
\includegraphics[width=9cm]{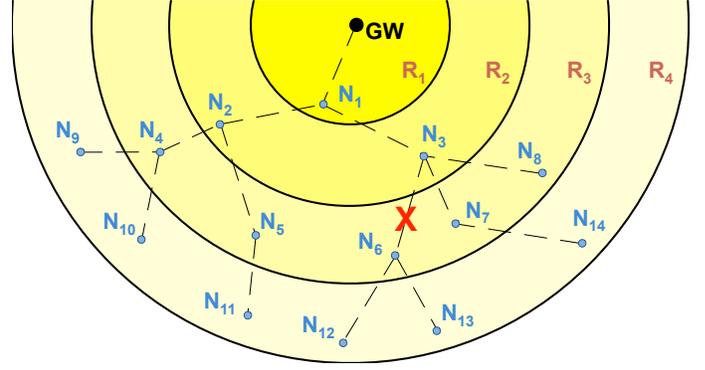}
\caption{Network topology of the multi-hop LPWAN from Figure \ref{fig:txwindow}, with a gateway (GW) and 14 stations ($N_{1}-N_{14}$) deployed in 4 rings ($R_{1}-R_{4}$).}
\label{fig:poison1}
\end{figure}

\begin{figure}
\centering
\includegraphics[width=9cm]{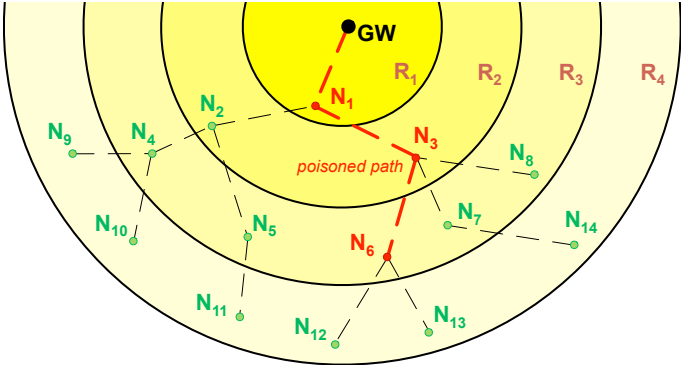}
\caption{State of the network from Figure \ref{fig:poison1} after the corresponding e2e ACK. Note the \textit{poisoned path} passing through nodes $N_{6}$, $N_{3}$, and $N_{1}$. Together with the GW, these nodes (colored in red) stay awake during the second transmission window. The rest of nodes (colored in green) go to sleep as they are not involved in the new transmission process.}
\label{fig:poison2}
\end{figure}	

In Figure \ref{fig:poison1}, node $N_{3}$ activates its poison flag after not receiving data from its child $N_{6}$. In its way to the GW, a data packet from $N_{3}$ poisons its next hop: $N_{1}$. Therefore, nodes $N_{6}$, $N_{3}$, and $N_{1}$ form a \textit{poisoned path}, as shown in Figure \ref{fig:poison2}.

\subsubsection{Transmission windows}

A number of transmission windows $(w)$ with their corresponding e2e ACK are included in each uplink data transmission phase to ensure correct packet reception. Within these windows, not all STAs remain awake, but only the ones directly involved in the transmission process. Before the start of a new transmission window, STAs evaluate whether they shall stay awake or go to sleep. 

\begin{figure}
\centering
\includegraphics[width=8.5cm]{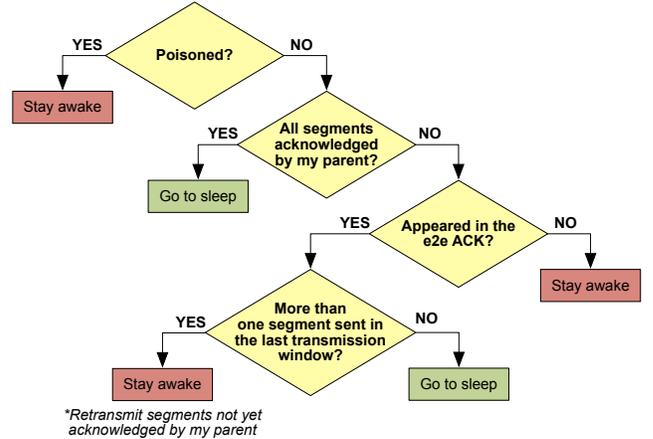}
\caption{STA's decision flowchart to stay awake or go to sleep before the start of a new transmission window.}
\label{fig:flowchart}
\end{figure}

This decision takes into account if the STA has been previously \textit{poisoned} by one of its children as well as several other conditions according to the decision flowchart from Figure \ref{fig:flowchart}. Whenever an STA decides to go to sleep, it will remain in this state until the next \textit{primary beacon}.

\subsubsection{Distributed caching}

Due to the structure of multi-hop networks, lost packets cause expensive retransmissions along every hop of the path between the sender and the receiver \cite{dunkels2004distributed}. To alleviate this problem, a distributed caching system is used in HARE, so that parents acknowledge the correct reception of packets from children and cache their data until it is properly received in the GW. 


As it can be seen in Figure \ref{fig:poison2}, nodes $N_{12}$ and $N_{13}$ can go to sleep after the first transmission window, because their data packets have been acknowledged by node $N_{6}$, which will cache them in memory together with its own data to be sent in the next transmission window.


\section{Testbed} \label{testbed}

Contiki 3.0 OS \cite{dunkels2004contiki} was the selected RTOS to validate the HARE protocol stack, mainly due to its ability to easily execute multiple processes concurrently and its powerful COOJA network simulator \cite{osterlind2006cross}. Apart from simulations, two different real platforms\footnote{See Contiki main web page (http://www.contiki-os.org/) for a comprehensive table of hardware compatible with Contiki 3.0 OS} were used for preliminary testing and operational validation: MEMSIC\texttrademark \:TelosB 2.4 GHz nodes \cite{memsic2016telosb} and Zolertia\texttrademark \:RE-Mote 868 MHz nodes \cite{zolertia2016remote}, whose main features are depicted in Table \ref{hardware}.

HARE protocol stack has been programmed as an additional module for Contiki 3.0 OS which interacts with the already available upper communication layers of the system (MAC and Network), regardless the employed hardware. Specific interactions of HARE with PHY layers of the aforementioned hardware were separately programmed.

\begin{table}[hbp]
\centering
\caption{Main features of the hardware employed in the HARE operational validation}
\label{hardware}
\begin{tabular}{|l|l|l|}
\hline
\textbf{Platform} & \textbf{MEMSIC TelosB} & \textbf{Zolertia RE-Mote} \\ \hline
Microprocessor    & TI MSP430              & ARM Cortex-M3             \\ \hline
Radio Module      & TI CC2420              & TI CC1200                 \\ \hline
Frequency Band    & 2.4 GHz                & 868/915 MHz               \\ \hline
\end{tabular}
\end{table}

Performance evaluation of HARE protocol stack was performed in a testbed located on the 2nd floor, right wing of the Tanger building at UPF facilities\footnote{UPF communication campus main website:\\ https://www.upf.edu/campus/en/comunicacio/tanger.html}. The testbed consisted of 13 Zolertia\texttrademark \: RE-Mote nodes (one of them acting as a gateway and connected to a PC) running the HARE protocol stack.

\begin{figure*}[!htb]
\centering
\includegraphics[width=14cm]{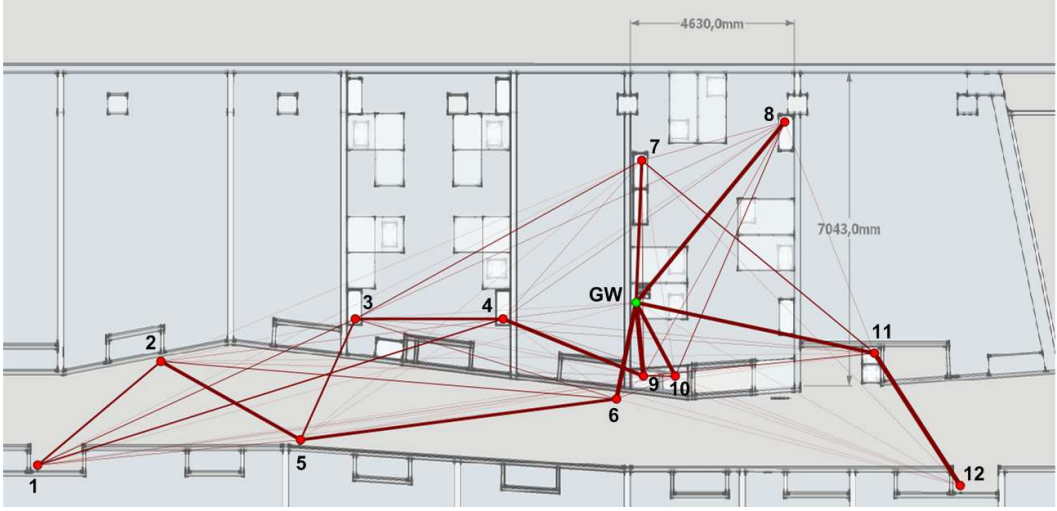}
\caption{Nodes' placement at UPF facilities and association diagram when each STA admits up to 5 children.}
\label{fig:association}
\end{figure*}

All tests were executed considering no mobility and with the same STAs' placement (see Figure \ref{fig:association}). All STAs were powered by an 800 mAh battery except the gateway, which was permanently powered by the PC. Results were directly obtained from the GW, or thanks to the \textit{statistics messages} periodically sent by STAs. These messages contain information about different metrics such as the number of packets sent and acknowledged, RTT delays, as well as power profiles of microprocessor and radio module.

The calculation of total energy consumption $(E_{\text{T}})$ is based on these two power profiles: $E_{\mu \text{P}}$ and $E_{\text{RADIO}}$, for the microprocessor and the radio module, respectively, as shown in Equation \eqref{energy_computing}. $V_{\text{DD}}$ is the supply voltage, while $t$ and $I$ are, respectively, the time and the current corresponding to the operational states of the microprocessor and the radio module of the employed hardware, whose values are summarized in Table \ref{values}. Notice that the $I_{\text{TX}}$ value of the radio module depends on the transmission power level.
\begin{eqnarray}
E_{\text{T}} &=& E_{\mu \text{P}} + E_{\text{RADIO}} \nonumber \\
E_{\mu \text{P}} &=& V_{\text{DD}} \cdot \left(t_{\text{CPU}} \cdot I_{\text{CPU}} + t_{\text{LPM}} \cdot I_{\text{LPM}}\right) \nonumber \\
E_{\text{RADIO}} &=& V_{\text{DD}} \cdot \left(t_{\text{RX}} \cdot I_{\text{RX}} + t_{\text{TX}} \cdot I_{\text{TX}} + t_{\text{SL}} \cdot I_{\text{SL}}\right)
\label{energy_computing}
\end{eqnarray}

\begin{table}[tbp]
\centering
\caption{Current values of the Zolertia RE-Mote different operational states (from \cite{texas2015cc2538})}
\label{values}
\begin{tabular}{|l|c|c|}
\hline                          & \textbf{Operational state} & \textbf{Current} \\ \hline
\multirow{2}{*}{\begin{tabular}[c]{@{}l@{}}\textbf{Microprocessor} \\ ARM Cortex-M3\end{tabular}} & Processing (CPU) & $I_{\text{CPU}}=13 mA$\\ \cline{2-3} 
                                &   Low power mode (LPM)       &    $I_{\text{LPM}}=0.4 \mu A$   \\ \hline
\multirow{3}{*}{\begin{tabular}[c]{@{}l@{}}\textbf{Radio Module} \\ TI CC1200\end{tabular}}   &  Receiving (RX) & $I_{\text{RX}}=19 mA$      \\ \cline{2-3} 
                                &   Transmitting (TX) & $I_{\text{TX}}=39-61 mA$      \\ \cline{2-3} 
                                &   Sleeping (SL) & $I_{\text{SL}}=0.12 \mu A$       \\ \hline
\end{tabular}
\end{table}

In addition, different network configurations were applied. Firstly, two different MAC layers inherent to Contiki OS were tested: NULLMAC and X-MAC \cite{buettner2006x}. While NULLMAC maintains STAs continuously awake during \textit{active} periods, X-MAC combines the introduction of sleeping periods for receivers with the use of strobed preambles for senders. 

Secondly, and always over the same node deployment, two different network topologies were tested: single-hop and multi-hop. In the first case, all nodes were directly connected to the GW, while in the second case, STAs were free to establish their own routes to the GW with the single limitation of having 5 children per STA. 

\begin{table}[!b]
\centering
\caption{Definition of error configurations for the proposed testbed}
\label{packet_error}
\begin{tabular}{|c|c|c|}
\hline
\textbf{Error Config.} & \textbf{Data Error} & \textbf{ACK Error} \\ 
\hline
\textbf{$E_{0\//0}$} & 0\% & 0\%\\ 
\hline
\textbf{$E_{10\//5}$} & 10\% & 5\%\\ 
\hline
\textbf{$E_{20\//10}$} & 20\% & 10\%\\ 
\hline
\textbf{$E_{30\//15}$} & 30\% & 15\%\\ 
\hline
\end{tabular}
\end{table}

And thirdly, the whole system was altered with the arbitrarily introduction of a certain error probability when sending both \textit{application packets} and their corresponding ACKs (it is worth noting here that neither messages implied in the association process nor \textit{statistics packets} were affected by arbitrary generated errors). Errors were generated through a uniformly distributed random variable according to mean error values from Table \ref{packet_error}. Before sending a message, STAs computed this value and discarded messages accordingly. For this purpose, four different error configurations were defined.

The addressing system followed the Rime format \cite{dunkels2007rime} consisting of two 8-bit  numbers. Similarly to IP addressing, the use of netmasks leads to flexible subnetting configurations with up to ($2^{16}-2$) STAs. In our particular case, the first 8-bit number identified the network prefix shared by all devices, and the second one the host part, whose value for GWs was 0 and for STAs was selected from 1 to 255. 

All tests began with a \textit{network association primary beacon} in which all nodes tried to associate to the network. From then on, the GW emitted a new (\textit{network association} or \textit{data}) \textit{primary beacon} every $T_{p} = 3$ min. \textit{Data primary beacons} could ask STAs for a new \textit{application} or \textit{statistics packet}. In all our tests, \textit{application} and \textit{statistics packets} generated by STAs contained, respectively, 10 and 20 bytes of net information\footnote{Implementation of IEEE 802.15.4 in Contiki OS increases the minimum length of any transmitted packet up to 43 bytes after including headers and, if necessary, applying padding}.

\section{Results} \label{results}

\subsection{Association process}
To show the performance and the coherence of the proposed association process and its underlying routing, all STAs were forced to repeatedly renew every two \textit{primary beacons} ($N_{\text{pr}}=2$) their association to the network and compute their best parent according to \eqref{routing_weights} with the following parameters: $a_{1}=a_{2}=10$, $a_{3}=1$, and $a_{4}=5$. In addition, the number of children per STA was artificially limited to $5$ to guarantee multiple paths towards the GW. Interspersed \textit{data primary beacons} were used to check the reliability of routing paths and to allow not yet associated STAs to have another opportunity to join the network.

The selected underlying MAC for all STAs was X-MAC and no error was introduced in the network (i.e., $E_{0\//0}$ error configuration was used). Under these premises, and after 200 repetitions, an average number of $11.97$ STAs were associated to the network after the \textit{data primary beacon} of the given sequence (i.e., $99.75\%$ of success). As for the packet delivery ratio (PDR), it achieved $100\%$ in all the associated STAs.

Routing tables compiled by the GW were processed and adapted to graphical representation in Figure \ref{fig:association}, where line's thickness is proportional to link's frequency appearance. Preference of STAs for establishing paths with closer neighbours in their way to the GW becomes evident, just like the importance of \textit{clear paths} (i.e., without obstacles) such as the formed by the corridor walls. 

The limitation of 5 children can be clearly appreciated in STAs $\#6$, $\#8$, $\#9$, $\#10$, and $\#11$ being almost always directly connected to the GW in ring 1. The rest of STAs (principally $\#7$) could only access to that ring when circumstantially having better channel conditions than the aforementioned ones.

\subsection{Reliability}
Once all STAs are associated to the network and their paths to the GW properly established, the next goal is to analyze the reliability and the cost (in terms of energy consumption) of sending data. To do that, the GW was programmed to send 20 beacons with the following sequence: beacon $\#1$ was a \textit{network association primary beacon}, beacons $\#10$ and $\#20$ were \textit{data primary beacons} asking for \textit{statistics packets}; the rest of beacons were \textit{data primary beacons} asking for \textit{application packets}. To send their packets, STAs had 5 available transmission windows $(w=5)$. 

\begin{figure*}[tbp]
\centering
\subfigure[NULLMAC]{\includegraphics[width=90mm]{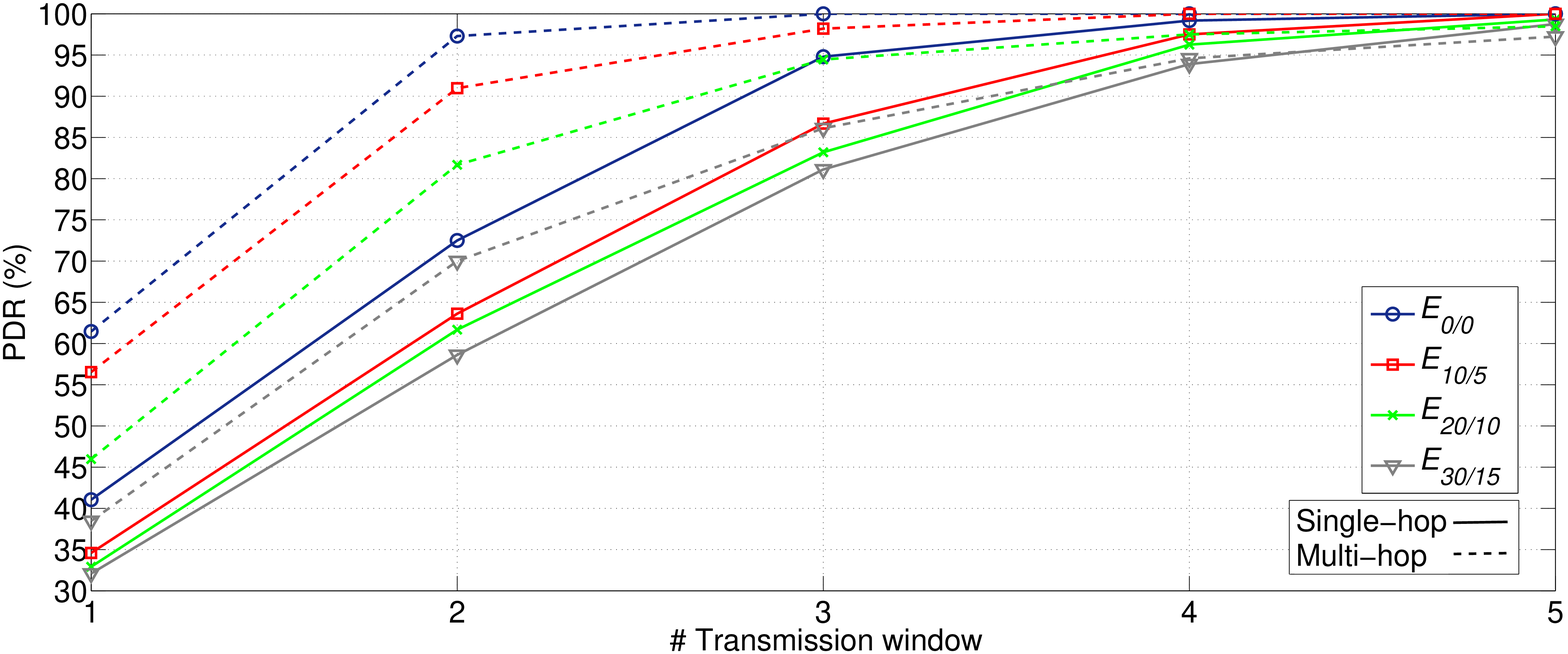}}
\subfigure[X-MAC]{\includegraphics[width=90mm]{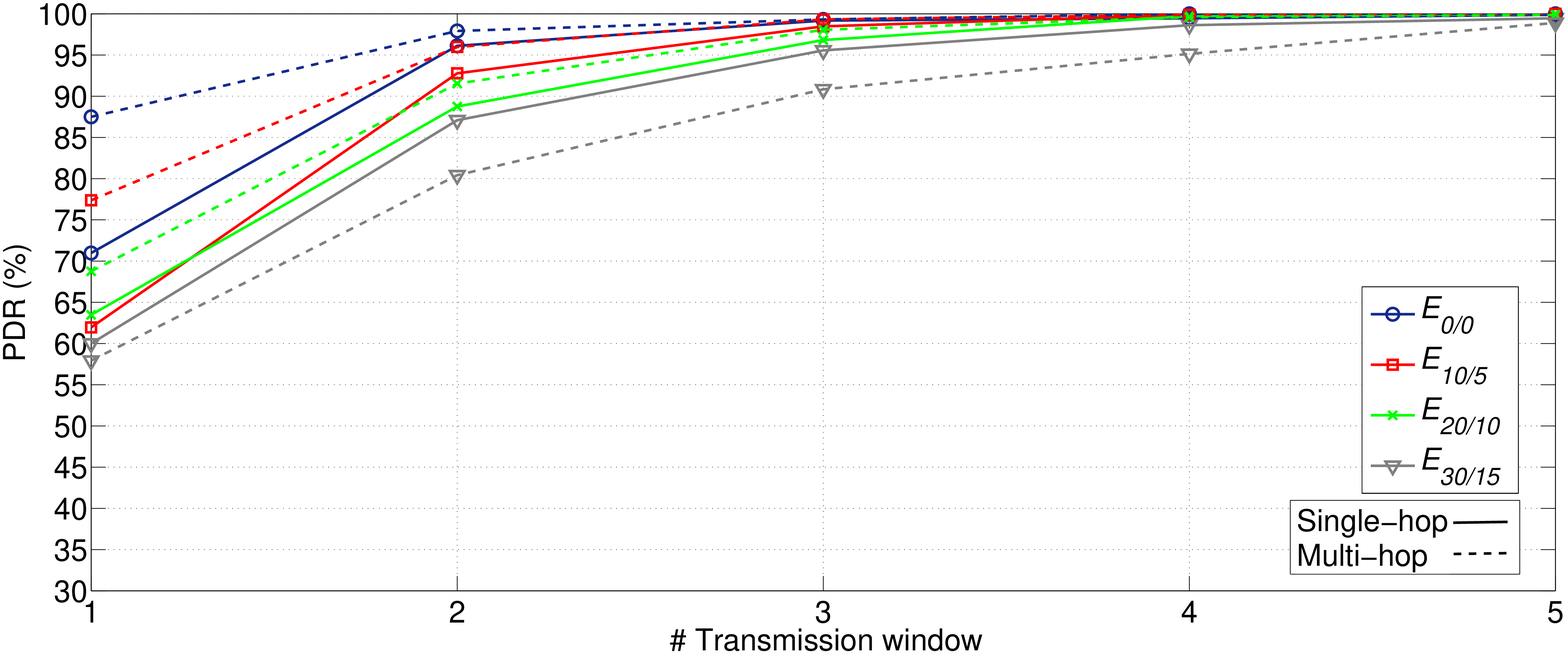}}
\caption{Packet delivery ratio in the proposed testbed for different MAC layers, network topologies, and error probabilities.} 
\label{fig:pdr}
\end{figure*}

The results with the obtained PDR in all these configurations are compiled in Figure \ref{fig:pdr}. After 5 transmission windows, PDR is in any configuration above 95\%, and it even achieves values above 90\% after 3 and 4 transmission windows when using X-MAC and NULLMAC, respectively. In this case, NULLMAC specially suffers from the effect of collisions, due to the backoff implementation\footnote{Main values of the NULLMAC CSMA/CA default backoff implementation in Contiki OS: minimum value of the backoff exponent ($macMinBE=0$), maximum value of the backoff exponent ($macMaxBE=4$), and maximum number of backoff attempts ($macMaxCSMABackoffs=5$).} and the higher number of concurrently active STAs compared to X-MAC. 

Another insight from obtained results is how multi-hop topology outperforms single-hop in all possible configurations except when using X-MAC with $E_{30\//15}$. Again, the inherent reduction of concurrently active STAs competing for the channel during the same time period (in this case, due to the allocation of STAs to different slots according to their ring) proves beneficial for system's reliability. 

\begin{figure*}[h!]
\centering
\includegraphics[width=10cm]{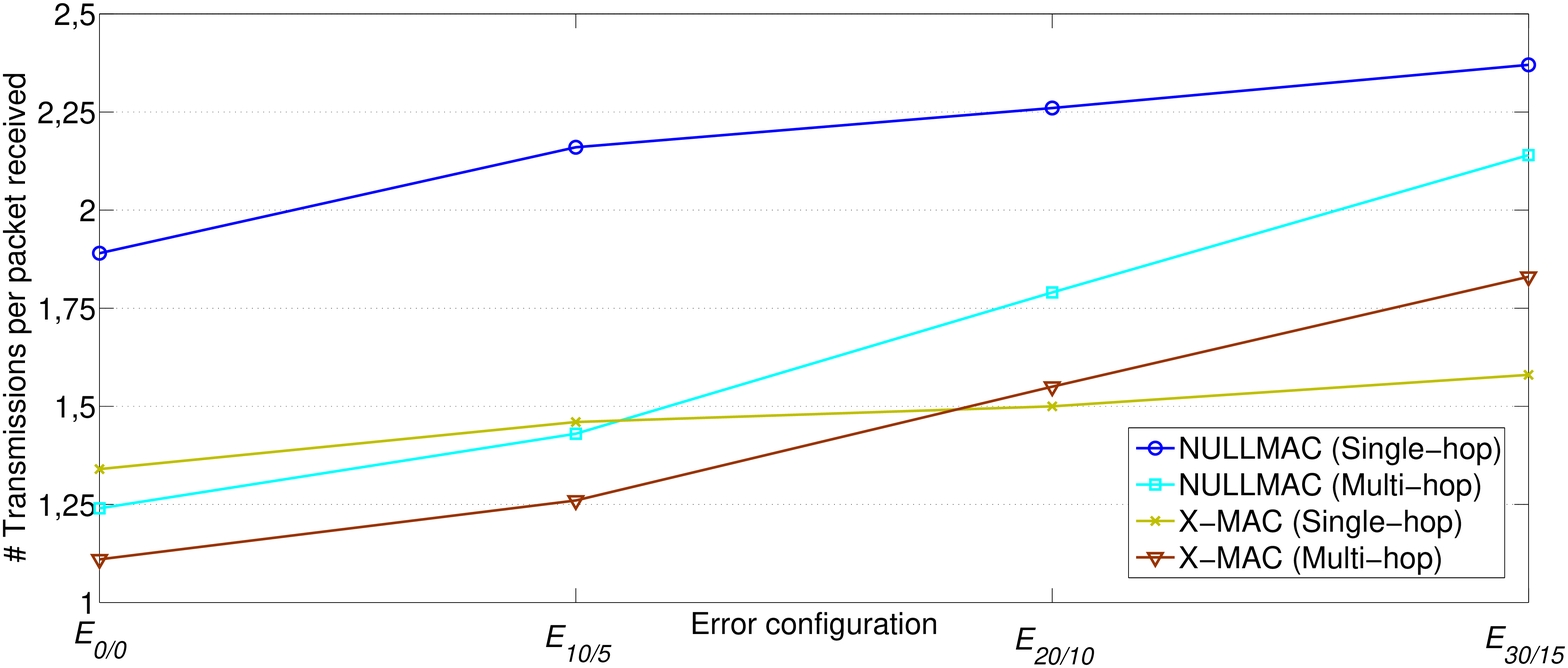}
\caption{Number of transmissions per packet received in the proposed testbed.}
\label{fig:trans}
\end{figure*}

The network's ability to properly deliver data packets to its destination was also analyzed by computing the quotient between the total number of packets sent by STAs and those properly received by the GW. As shown in Figure \ref{fig:trans}, multi-hop schemes still have better performance than single-hop in low-error configurations. On the contrary, in highly unfavorable channels, parents usually do not receive all their expected payloads at once, so that they tend to send several packets in successive transmission windows with only partial information.

\subsection{Energy consumption}

\begin{figure*}[htbp]
\centering
\includegraphics[width=10cm]{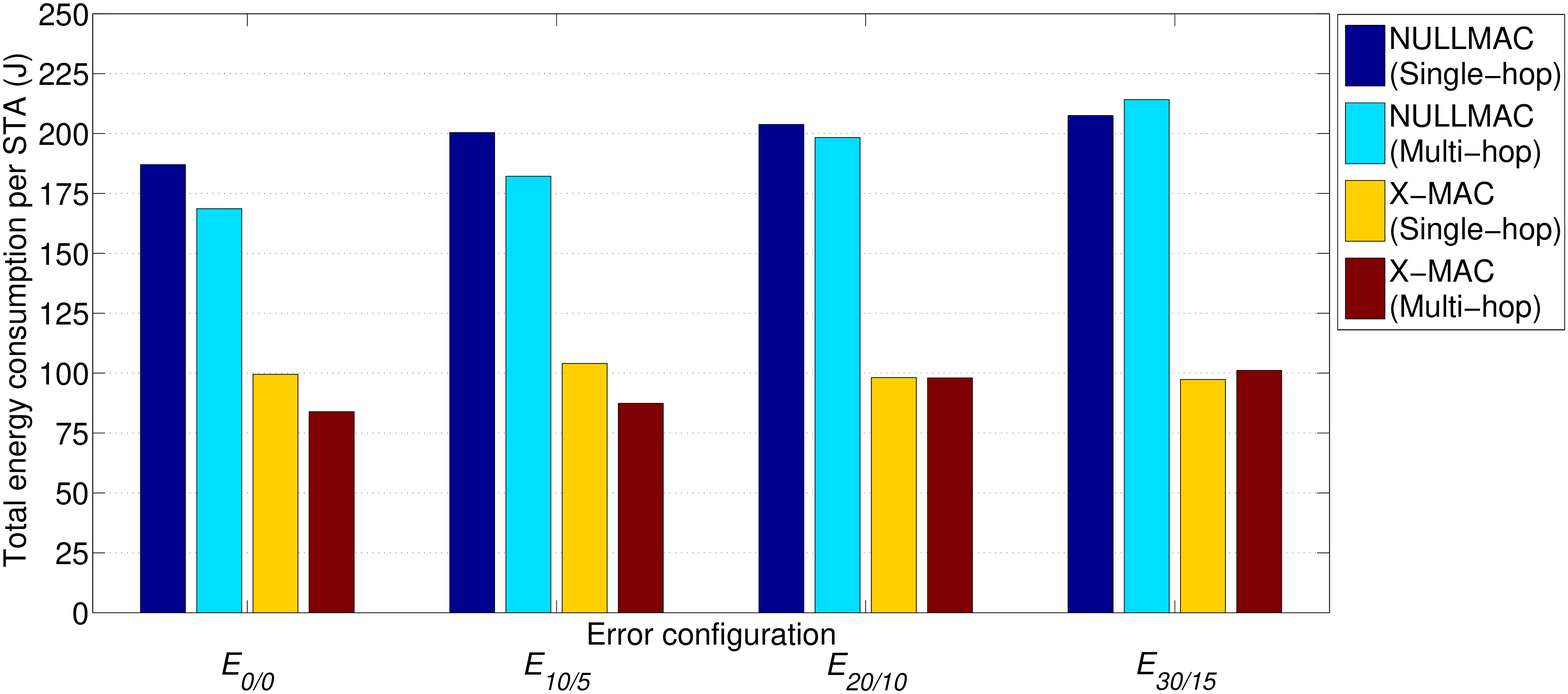}
\caption{Average total energy consumption per STA after 20 beacons when $T_{p}=3min.$}
\label{fig:energy}
\end{figure*}

The effect of this interdependence can also be observed in total energy consumption (Figure \ref{fig:energy}), computed after 20 transmitted beacons (i.e., a 1-hour test). Important savings (up to 15\%)\footnote{From previous studies \cite{barrachina2016multi}, we believe that in larger networks, these gains will be much higher.} can be achieved when using multi-hop schemes with respect to single-hop ones in low-error configurations $(E_{0\//0} - E_{20\//10})$ and similar or slightly worse values (less than 4\% of extra consumption) in $E_{30\//15}$. 

Time percentage of STAs' microprocessor in low power mode is, in all studied cases, above 97\% for X-MAC and 99\% for NULLMAC, due to the higher number of operations involved in the first case. However, the impact of radio module sleeping periods introduced by X-MAC layer reduces total energy consumption in up to 50\% with respect to NULLMAC. 
In this case, values of energy consumed per bit of payload delivered are confined between $50 - 65$ mJ\//bit for X-MAC, and $105 - 140$ mJ\//bit for NULLMAC.

As for the battery lifetime, Table \ref{lifetime} compiles the duration in days of the 800mAh battery included in the Zolertia RE-Mote for the current testbed with $T_{p}=3$ min, as well as two estimations with $T_{p}=1$ h and $T_{p}=4$ h. The temporal flexibility of the TDMA-based system employed in HARE allows this kind of extrapolations, by assuming that, in non-active time periods, both the microprocessor and the radio module remain asleep.

\begin{table}[]
\scriptsize
\centering
\caption{Average lifetime of an 800 mAh battery in the proposed testbed}
\label{lifetime}
\begin{tabular}{|c|c|c|c|c|c|}
\hline
\multicolumn{3}{|c|}{\multirow{2}{*}{}}                                        & \multicolumn{3}{c|}{\textbf{Battery lifetime (days)}} \\ \cline{4-6} 
\multicolumn{3}{|l|}{}                                                         & $T_{p}=3min$    & $T_{p}=1h$    & $T_{p}=4h$    \\ \hline
\multirow{8}{*}{\rotatebox{90}{\textbf{NULLMAC}}} & \multirow{4}{*}{\textbf{Single-hop}}                      & $E_{0\//0}$  &  $2.37$  & $47.35$   &  $187.87$  \\ \cline{3-6} 
                         &                                                  & $E_{10\//5}$ & $2.21$   & $44.19$   & $175.41$   \\ \cline{3-6} 
                         &                                                  & $E_{20\//10}$ &  $2.18$  & $43.46$   & $172.53$    \\ \cline{3-6} 
                         &                                                  & $E_{30\//15}$ & $2.14$   & $42.70$   & $169.53$   \\ \cline{2-6} 
                         & \multirow{4}{*}{\textbf{Multi-hop}}                       & $E_{0\//0}$ & $2.63$   &  $52.51$  & $208.17$   \\ \cline{3-6} 
                         &                                                  & $E_{10\//5}$ &  $2.44$  & $48.60$   & $192.79$   \\ \cline{3-6} 
                         &                                                  & $E_{20\//10}$ &  $2.24$  & $44.66$   & $177.27$   \\ \cline{3-6} 
                         &                                                  & $E_{30\//15}$ &  $2.07$  &  $41.35$  & $164.23$   \\ \hline
\multirow{8}{*}{\rotatebox{90}{\textbf{X-MAC}}} & \multirow{4}{*}{\textbf{Single-hop}}                      & $E_{0\//0}$  &  $4.46$  &  $88.75$  & $349.64$   \\ \cline{3-6} 
                         &                                                  & $E_{10\//5}$  &  $4.27$  & $85.00$   &  $335.07$    \\ \cline{3-6} 
                         &                                                  & $E_{20\//10}$  &  $4.52$  & $89.99$   &  $354.46$     \\ \cline{3-6} 
                         &                                                  & $E_{30\//15}$  & $4.56$   & $90.79$   &  $357.55$   \\ \cline{2-6} 
                         & \multirow{4}{*}{\textbf{Multi-hop}}                       & $E_{0\//0}$  & $5.29$   &  $105.17$  & $413.16$  \\ \cline{3-6} 
                         &                                                  & $E_{10\//5}$  &  $5.08$  &  $101.04$  & $397.21$     \\ \cline{3-6} 
                         &                                                  & $E_{20\//10}$  & $4.53$   &  $90.10$  & $354.85$   \\ \cline{3-6} 
                         &                                                  & $E_{30\//15}$  &  $4.39$  & $87.38$   &  $344.31$   \\ \hline
\end{tabular}
\end{table}

\subsection{Resilience against failures}

To prove the adaptability and resilience of the routing protocol implemented in HARE, the network was subjected to the deliberate shutdown of two of its STAs. In this way, the GW was programmed to send 50 beacons with the following sequence: beacon $\#1$ was a \textit{network association primary beacon}, beacons multiple of 10 were \textit{data primary beacons} asking for \textit{statistics packets}; the rest of beacons were \textit{data primary beacons} asking for \textit{application packets}. In addition, the \textit{disassociation mechanism} was programmed in the GW to remove an STA from the network if not receiving any data packet during one \textit{primary beacon} ($N_{\text{pd}}=1$).

\begin{figure}[H]
\centering
\subfigure[Logical network topology after the \textit{network association primary beacon} \label{fig:top_1}]{\includegraphics[width=42mm]{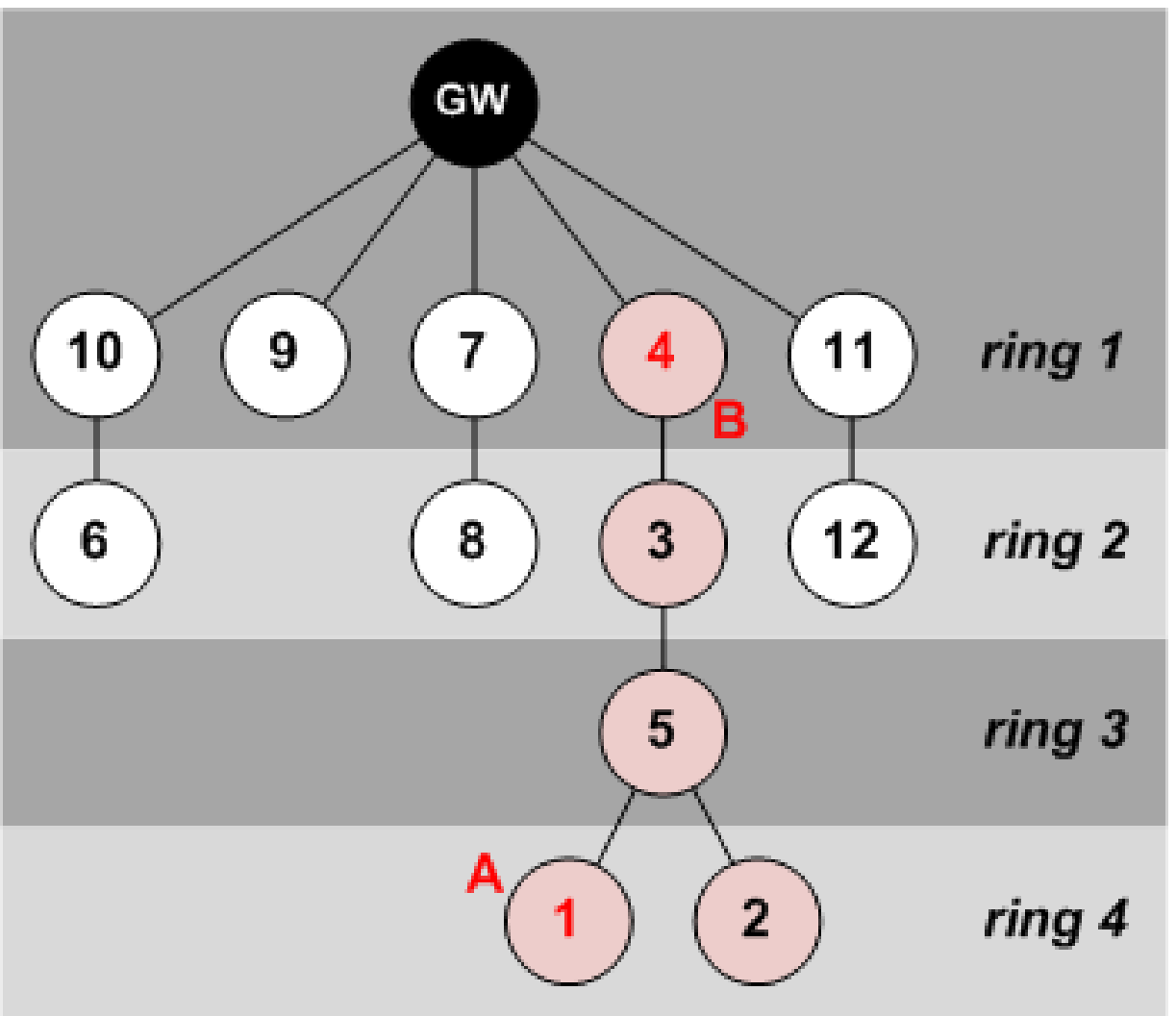}}
\subfigure[Logical network topology from beacon \#15 until beacon \#50 \label{fig:top_2}]{\includegraphics[width=42mm]{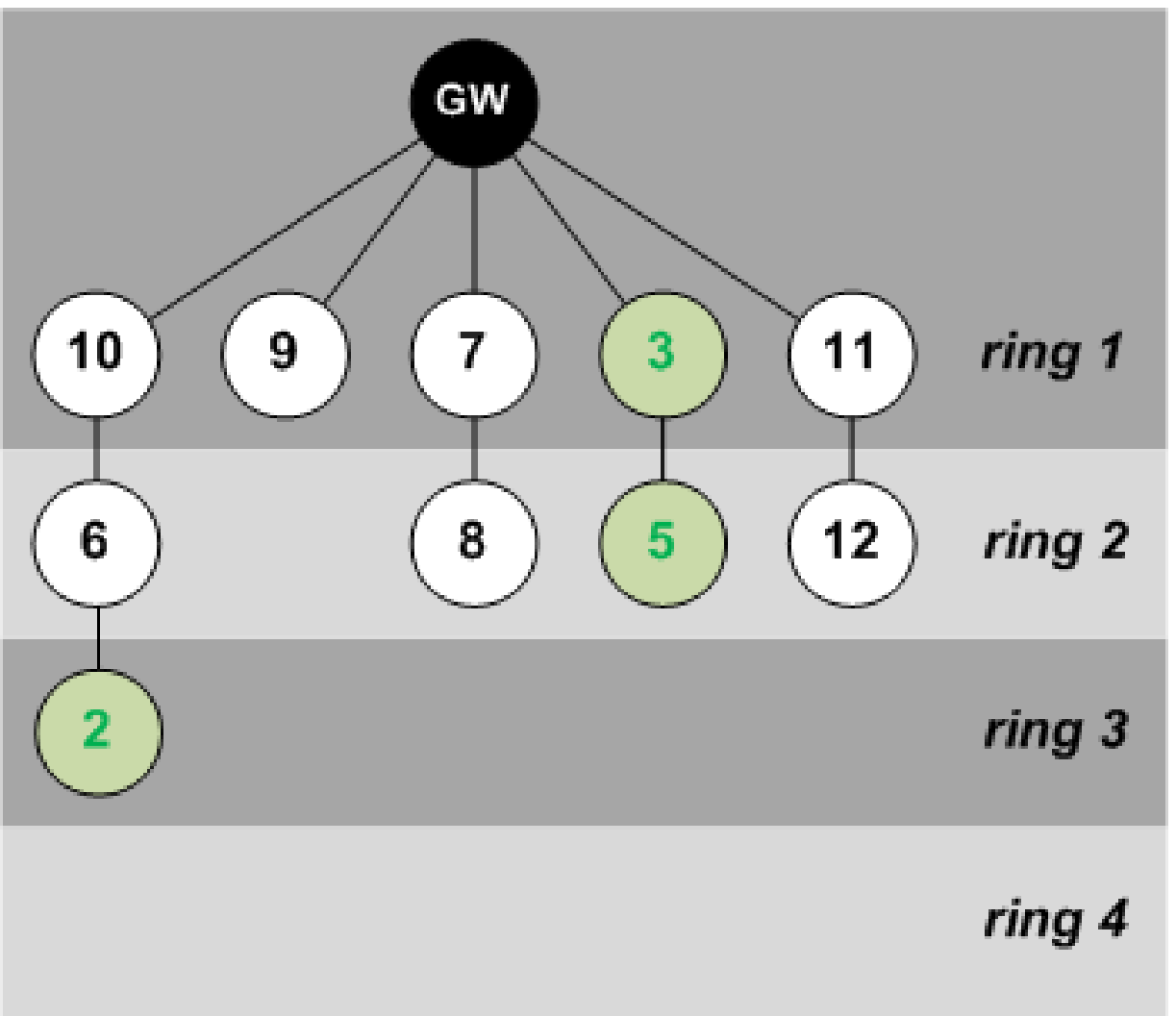}}
\caption{Network topology before and after shutdown of nodes \#1 and \#4.} \label{fig:topology}
\end{figure}

Once finished the initial \textit{network association mechanism}, the network was organized in four rings, as shown in Figure \ref{fig:top_1}. After beacon \#4 \textbf{(A)}, STA \#1 was switched off, but it did not imply further problems to the network, as this STA did not have any children. However, after beacon \#12 \textbf{(B)}, STA \#4 was also switched off, and it forced the network to reconfigure itself. The path to the GW of STAs \#2, \#3 and \#5 was broken, and they had to look for a new route by using the \textit{STA association mechanism} of successive \textit{data primary beacons}. After beacon \#15, all active STAs (i.e., all of them except \#1 and \#4, which remain off) had a path to the GW and the network was again stable (see Figure \ref{fig:top_2}).

This test was also useful to analyze the performance of the proposed \textit{power regulation mechanism} when setting it with $\text{RSSI}_{min}=-110$ dBm and $\text{RSSI}_{max}=-100$ dBm. It is worth noting here that Zolertia RE-Mote devices use up to 31 different power levels (from $-16$ dBm to $14$ dBm with steps of 1 dB \cite{texas2013an125}) and are programmed by default with the maximum transmission power level.

\begin{figure*}[t!!]
\centering
\includegraphics[width=17cm]{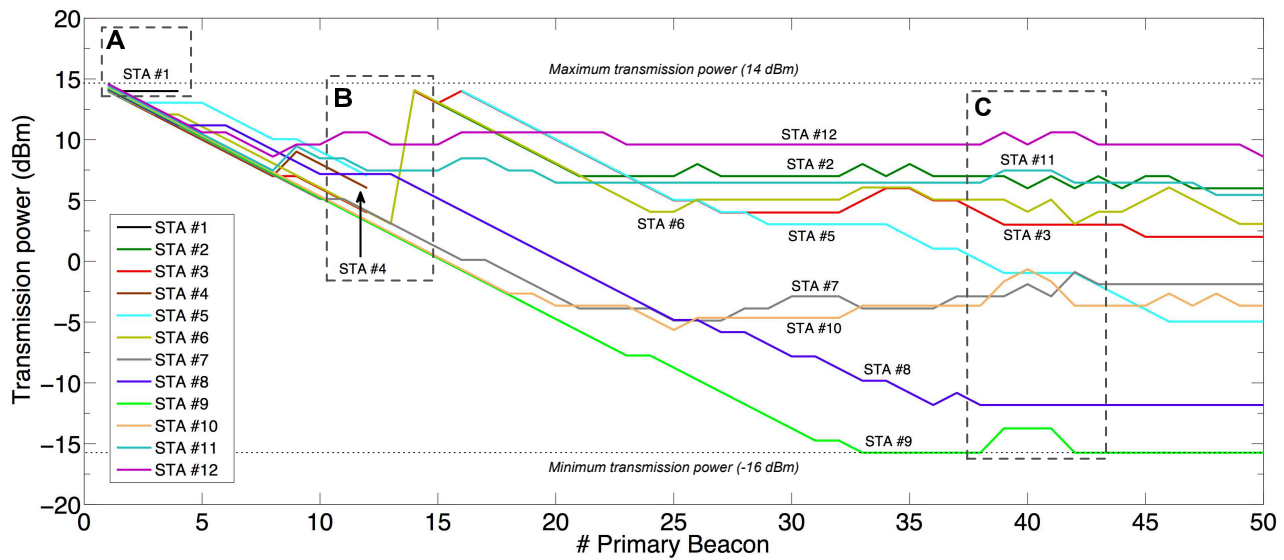}
\caption{Temporal evolution of STAs' transmission power level.}
\label{fig:resilience}
\end{figure*}

Figure \ref{fig:resilience} shows the clear reduction of transmission power in most of the analyzed STAs during 50 \textit{primary beacons}, being the most significant examples STAs \#7, \#8, \#9 and \#10; the nearest ones to the GW. This fact results in a lower energy consumption, as $I_{TX}=61$ mA when transmitting at 14 dBm, but almost half ($I_{TX}=39$ mA) when doing it at -16 dBm. 

The effects of switching off nodes are also visible in the transmission power, as shown in \textbf{(A)} and \textbf{(B)} from Figure \ref{fig:resilience}. While STA \#1 in \textbf{(A)} simply stopped working, nodes involved in the shutdown of STA \#4 in \textbf{(B)} experienced notable changes. Thus, STAs \#2, \#3 and \#5 \textit{disappeared} along with the shutdown of STA \#4. However, they became associated again between beacons \#13 and \#15 with maximum transmission power. For its part, when STA \#6 became parent of STA \#2, it set the maximum power level to establish connection with its new child.

Lastly, the \textit{power regulation mechanism} proved its good performance against channel alterations as shown in area \textbf{(C)}. In this case, and due to the test execution on a real scenario, the presence of people in the floor corridor may have disturbed channel conditions. To overcome this situation, some STAs (\#9, \#10, \#11 and \#12)  selected temporarily greater transmission power levels that were reestablished once finished the detected channel issues.

\section{Conclusions} \label{conclusions}

Single-hop topologies have become the \textit{de facto} system to transmit data in current LPWANs mainly due to the need for network simplicity and robustness, and the fear of consuming too much energy in processing tasks and/or maintaining complex routing mechanisms. However, the HARE protocol stack presented in the current article proves the suitability of alternative uplink multi-hop communication approaches in LPWANs. Distributed among three OSI layers (MAC, network and transport), the multiple mechanisms contained in HARE ensure network reliability and resilience against failures in uplink transmissions while keeping low energy consumption. 

Results from a real testbed show uplink PDR values above 95\% for all considered configurations, with faster achievement of this level when using multi-hop topologies with multiple transmission windows. In addition, multi-hop topologies outperform single-hop ones in terms of energy consumption in the considered non error-prone scenarios, with up to 15\% improvement (which could be even much higher in larger networks) and values as low as $50 mJ/bit$ when employing an underlying duty-cycled MAC layer. Similarly, network auto-configuration and resilience have been successfully put to the test after forcing the shutdown of some network STAs.

In the near future, LPWANs are foreseen to occupy a central role in applications requiring to interconnect low-bandwidth devices, focusing on range and power efficiency. While range coverage is mostly an issue from the physical layer, future challenges regarding power efficiency will surely encompass the coordination of different layers and even the inclusion of novel cross-layer mechanisms. 

\section*{Acknowledgments}

This work was partially supported by the Catalan government through the project SGR-2014-1173 and also received funding from the European Union’s Seventh Framework Programme for research, technological development and demonstration under grant agreement no 605073 (ENTOMATIC).

\bibliographystyle{IEEEtran}
\bibliography{Bib}

\end{document}